\documentclass[12pt]{article}

\usepackage[T1]{fontenc}
\usepackage[utf8]{inputenc}
\usepackage{lmodern}
\usepackage{amsmath,amssymb,amsfonts,amsthm}
\usepackage{graphicx}
\usepackage{bm}
\usepackage{geometry}
\usepackage{booktabs}
\usepackage{multirow}
\usepackage{array}
\usepackage{slashed}
\usepackage[numbers,sort&compress]{natbib}
\usepackage{hyperref}
\usepackage{xcolor}
\usepackage{caption}
\usepackage{enumitem}
\usepackage{microtype}
\usepackage{float}

\hypersetup{
  colorlinks=true,
  linkcolor=blue,
  citecolor=blue,
  urlcolor=blue
}

\setlist[itemize]{leftmargin=2em}
\setlist[enumerate]{leftmargin=2em}

\newcommand{\Disc}{\operatorname{Disc}}

\title{\boldmath Target-Mass Corrections in the OPE Sum-Rule Approach
to Quarkonium--Nucleon Interactions with Global-Fit PDFs:
an $x$-Resolved Analysis}

\author{Arkadiy I.~Syamtomov\\[0.5em]
\small Bogolyubov Institute for Theoretical Physics,\\
\small National Academy of Sciences of Ukraine,\\
\small Kiev, Ukraine\\[0.3em]
\small \texttt{arkady.syamtomov@gmail.com}}

\date{}

\begin{document}

\maketitle

\begin{abstract}
We derive an exact all-orders treatment of target-mass corrections (TMCs) within the leading-twist Coulombic OPE for the absorptive quarkonium--nucleon spectral baseline. Resumming the complete trace series before analytic continuation yields a closed spectral kernel, a shifted-convolution expression, and a moment representation that preserves the on-shell incoming domain $0<y\leq1$, with $y=m_N/\lambda$ and $\lambda$ the nucleon energy in the quarkonium rest frame. Using native ABMP16, MSHT20, CT18, and NNPDF4.0 grids, we find at $Q=10$~GeV and $\epsilon_0=0.16$~GeV that the TMC/no-TMC moment ratio decreases from about $0.85$ at $n=4$ to $0.42$--$0.48$ at $n=12$. At the first tabulated energies above the incoming two-body endpoint, the leading-twist cross-section ratio is strongly suppressed but rapidly approaches unity with increasing energy. The strict endpoint has an integrable inverse-velocity behaviour, whose hadronic-threshold interpretation lies outside the leading-twist construction.
\end{abstract}

\section{Introduction}

The interaction of heavy quarkonia with hadronic matter remains an important
theoretical probe of QCD dynamics at the interface of perturbative and
nonperturbative scales. Since heavy quarkonium states such as $J/\psi$ and
$\Upsilon$ are compact compared to ordinary hadrons, their interaction with
external gluonic fields can, at least in the short-distance picture, be treated
by means of the operator product expansion (OPE) and the multipole expansion.
In this formulation the short-distance dynamics of the heavy $Q\bar Q$ state is
encoded in Wilson coefficients, while the long-distance structure of the hadronic
target is described by matrix elements of local gluonic operators
\cite{Peskin:1979va,Bhanot:1979vb}.

In the sum-rule formulation, the forward quarkonium--nucleon amplitude is
expressed through moments of the gluon distribution in the nucleon,
\begin{equation}
A_n(Q)=\int_0^1 dx\,x^{n-2}g(x,Q),
\end{equation}
which enter as matrix elements of twist-2 gluonic operators of definite spin
\cite{Peskin:1979va,Bhanot:1979vb}.
The resulting sum rules relate Mellin moments of the gluon PDF to moments of
the forward quarkonium--nucleon scattering amplitude
\cite{KharzeevSatz:1994,Kharzeev:1996yx}.
The absorptive inelastic $\Phi N$ cross section is reconstructed from the
discontinuity of the resummed forward amplitude through the optical theorem.

An important refinement of this approach is the inclusion of finite target-mass
effects. In the original simplified treatment of the short-distance QCD
sum-rule approach~\cite{KharzeevSatz:1994}, trace terms in the target matrix
elements were neglected. They were subsequently restored
\cite{Kharzeev:1996yx}, where their fixed-spin sum was expressed through
a generalized hypergeometric weight. The present work revisits this construction
at the level of the fully resummed amplitude and shows that enforcing the
on-shell incoming domain $0<y\leq1$ prevents the on-shell moment from
factorising into that fixed-spin weight alone. Neglect of the trace terms is
justified only when $\tau = m_N^2/(4\epsilon_0^2) \ll 1$. The dimensionless parameter $\tau$ compares the
target-mass scale with the Coulombic binding-energy scale of the quarkonium state;
in the TMC kernel it enters through the combination $\tau x^2$, which determines
where finite-target-mass effects become numerically important. For $J/\psi$, $\tau\approx 8.6$, so this ratio is far from small.  A
consistent treatment must therefore retain the complete traceless tensor
structure, resum the resulting all-orders trace series at the amplitude level,
and only then perform the analytic continuation. This ordering is essential because,
with $\kappa\equiv m_N/\epsilon_0$ and $z=\kappa x/y$, the on-shell incoming
range $0<y\leq1$ implies $z\geq\kappa x$, while the partonic absorptive cut
requires $z\geq1$. The resulting spectral boundary is therefore
$z_{\min}(x)=\max(1,\kappa x)$, so the moment cannot be represented by a
simple factorised weight. Importantly, these target-mass corrections remain within
the twist-2 sector; they are kinematic consequences of trace subtraction and do not
introduce new higher-twist operators \cite{Kharzeev:1996yx}.

The same short-distance OPE formulation was subsequently used to connect the
absorptive $J/\psi N$ amplitude to photoproduction through vector-meson
dominance and a dispersion relation \cite{Kharzeev:1999}.  Later leading-twist
studies developed the quarkonium--hadron dissociation cross section directly
from gluon distributions \cite{Arleo:2001mp}, while recent near-threshold
photoproduction work has focused on gluonic gravitational form factors and the
proton mass distribution \cite{Kharzeev:2021qkd,Hatta:2019lxo,Boussarie:2020vmu,Mamo:2022eui}. Recent measurements are reported in \cite{GlueX2023,HallC2023}, while dispersive amplitude analyses are presented in \cite{Gryniuk:2016mpk}.
The present article addresses a narrower and complementary problem: the exact
kinematic target-mass structure of the twist-2 absorptive $\Phi N$ amplitude,
resolved in the gluon momentum fraction and evaluated with modern PDFs.

The present work is motivated by the fact that the current global-fit PDF
landscape differs substantially from that available in the 1990s. Global sets
such as ABMP16, MSHT20, CT18 and NNPDF4.0 incorporate a broader body of
DIS, collider and electroweak data and provide improved control over the gluon
distribution in the small-, intermediate- and large-$x$ regions
\cite{ABMP16,MSHT20,CT18,NNPDF40main}. This makes it natural to revisit the
quarkonium sum-rule calculation with modern input.

The central question therefore concerns the finite-energy consequences rather than the trace algebra alone: how strongly can the finite nucleon mass modify the $J/\psi N$ sum rules and the absorptive cross section, and over what energy range does the effect remain visible? We derive the trace-resummed spectral kernel generated by the gluon distribution and use it to obtain the on-shell moments and the finite-energy cross section. In this construction, the partonic momentum fraction remains correlated with the on-shell-domain boundary, which is particularly important near the incoming two-body endpoint.

The numerical analysis uses native LHAPDF grids \cite{LHAPDF6} for ABMP16,
MSHT20, CT18 and NNPDF4.0.  Their complete Hessian or replica ensembles are
propagated to the moment ratios and to
$\mathcal{R}_{\rm TMC}(\sqrt{s})$, while analytic five-parameter fits are
used only as auxiliary smooth representations and are documented in
\ref{app:pdf_params}.

The paper is organised as follows. Section~\ref{sec:formalism} introduces the
OPE formalism, derives the exact all-trace-order spectral kernel and physical moment,
and specifies the native-LHAPDF inputs and $x$-resolved diagnostics.
Section~\ref{sec:results} presents the exact direct-convolution construction
and the main moment-level and spectral-baseline results.
Section~\ref{sec:sensitivities} discusses scale dependence, binding-energy
sensitivity, numerical validation, the relation to photoproduction, and
higher-twist limitations. Section~\ref{sec:conclusions} summarises the
conclusions. The appendices document the auxiliary analytic PDF fits, the infrared
sensitivity of the excluded $n=2$ moment, the fixed-trace-order expansion used
to check the resummed kernel, and the derivation connecting the extended-domain
moment to the hypergeometric sum rule.

\section{OPE/TMC formalism and $x$-resolved moments}
\label{sec:formalism}

The object of the OPE analysis is the forward amplitude
$\Phi(q)N(p)\to\Phi(q)N(p)$, whose absorptive part determines the inelastic
$\Phi N$ cross section through the optical theorem. For a sufficiently heavy
quarkonium state, the $Q\bar Q$ pair is nonrelativistic and compact, with a
Coulombic Bohr radius $a_0$ \cite{Peskin:1979va,Bhanot:1979vb}. If this radius
is small compared with the hadronic scale, the forward amplitude can be expanded
in local gluonic operators, with the nucleon structure encoded in twist-2 matrix
elements related to Mellin moments of the gluon PDF:
\begin{equation}
\mathcal M_{\Phi N}
=
\sum_{n=2,4,\ldots} C_n\,\langle O_n\rangle,
\label{eq:OPE_amplitude}
\end{equation}
where the sum runs over even $n$ by charge-conjugation symmetry. In the
spin-averaged case, the relevant operator is the contraction of the quarkonium
four-momentum $K^{\mu}$ with the totally symmetric traceless twist-2 gluonic
operator $\theta_G^{\mu_1\cdots\mu_n}$ of rank $n$
\cite{KharzeevSatz:1994,Kharzeev:1996yx}:
\begin{equation}
O_n=
\frac{1}{M_\Phi^n}
K^{\mu_1}\cdots K^{\mu_n}\,
\theta_{G,\mu_1\cdots\mu_n}.
\end{equation}
Its nucleon matrix element takes the form
\begin{equation}
\langle p|\theta_{G,\mu_1\cdots\mu_n}|p\rangle
=
A_n(Q)\,
\bigl(
p_{\mu_1}\cdots p_{\mu_n}
-\text{traces}
\bigr),
\end{equation}
with $p^\mu$ the nucleon four-momentum. The coefficient $A_n(Q)$ is the
corresponding Mellin moment of the gluon distribution; its explicit PDF
representation is given below.

For a Coulombic $1S$ quarkonium state with Bohr radius $a_0=4/(3m_Q\alpha_s)$
and binding-energy parameter $\epsilon_0$, the Wilson coefficients derived in the
original short-distance quarkonium analysis \cite{Peskin:1979va,Bhanot:1979vb}
can be written as
\begin{equation}
C_n=a_0^3 \epsilon_0^{\,2-n} d_n,
\end{equation}
where $d_n$ is the dimensionless Wilson-coefficient factor generated by the
Coulombic quarkonium matrix element,
\begin{equation}
d_n=
\left(\frac{32}{N_c}\right)^2
\sqrt{\pi}\,
\frac{\Gamma\!\left(n+\frac52\right)}{\Gamma(n+5)}.
\end{equation}
The parameters entering the Coulombic coefficients are not refitted in the
present work. We use the standard phenomenological input of the original
quarkonium sum-rule applications, $m_Q=1.5$~GeV, $\alpha_s=0.3$, and
$\epsilon_0=0.16$~GeV \cite{Kharzeev:1996yx,Kharzeev:1999}. This choice keeps
the quarkonium part of the calculation fixed while the analysis focuses on the
propagation of modern gluon PDFs and target-mass corrections.  In this
phenomenological treatment, $\epsilon_0$ is retained as an independent
binding-energy input; it is not derived from the simultaneously adopted values
of $m_Q$ and $\alpha_s$.  The short-distance normalization is nevertheless kept
in the conventional Coulombic form used in \cite{Kharzeev:1996yx,Kharzeev:1999},
as specified below. Since $\epsilon_0$ also controls the TMC parameter
$\tau=m_N^2/(4\epsilon_0^2)$, its numerical impact is examined separately in
Sec.~\ref{sec:sensitivities}. The fixed inputs used in the central calculation
are listed in Table~\ref{tab:input_params}.

\begin{table}[h]
\centering
\caption{Fixed input parameters, conventional normalization, and derived kinematic quantities used in the central calculation.}
\label{tab:input_params}
\resizebox{\linewidth}{!}{%
\small
\begin{tabular}{llll}
\toprule
Parameter & Symbol & Value & Source or definition \\
\midrule
Nucleon mass        & $m_N$       & $0.938$~GeV  & \cite{PDG2024} \\
$J/\psi$ mass       & $M_{J/\psi}$& $3.097$~GeV  & \cite{PDG2024} \\
Charm quark mass    & $m_Q$       & $1.5$~GeV    & \cite{Kharzeev:1999} \\
Strong coupling     & $\alpha_s$  & $0.3$        & \cite{Kharzeev:1999} \\
Number of colours   & $N_c$       & $3$          & QCD \\
Binding-energy scale& $\epsilon_0$& $0.16$~GeV   & phenomenological input; \cite{Kharzeev:1999} \\
Reference PDF scale & $Q$         & $10$~GeV     & central; varied in Sec.~\ref{sec:sensitivities} \\
\midrule
Coulombic Bohr radius & $a_0=4/(3m_Q\alpha_s)$ & $2.963$~GeV$^{-1}$ & derived \\
Target-mass parameter & $\tau=m_N^2/(4\epsilon_0^2)$ & $8.592$ & derived \\
Kinematic ratio & $\kappa=m_N/\epsilon_0$ & $5.8625$ & derived \\
On-shell-domain boundary & $x_\star=1/\kappa$ & $0.1706$ & derived \\
Cross-section normalization & $K_\sigma=4/(3\alpha_s m_Q^2)$ & $1.9753$~GeV$^{-2}$ & conventional Coulombic normalization; \cite{Kharzeev:1996yx,Kharzeev:1999} \\
Incoming two-body endpoint & $\sqrt{s_{\rm ext}}=M_{J/\psi}+m_N$ & $4.035$~GeV & derived \\
\bottomrule
\end{tabular}}
\end{table}

\label{sec:sumrules}

The Wilson coefficients $C_n$ specify the short-distance quarkonium side of
the OPE. The corresponding long-distance nucleon input is the coefficient
$A_n(Q)$ in Eq.~\eqref{eq:OPE_amplitude}. In terms of the gluon PDF, it is
given by
\begin{equation}
A_n(Q)=\int_0^1 dx\,x^{n-2}g(x,Q).
\end{equation}
Thus each spin-$n$ operator probes a different power-weighted region of the
gluon distribution: lower moments are more sensitive to small $x$, while
higher moments progressively emphasize intermediate and large $x$.

At small $x$, the modern NNLO gluon distributions used here behave approximately as
$xg(x,Q)\propto x^{\alpha_g}$, where $\alpha_g$ denotes an effective small-$x$
exponent and is approximately $-0.3$ at the reference scale $Q=10$~GeV.
Consequently, $g(x,Q)\propto x^{\alpha_g-1}$, and the lowest moment
$A_2=\int_0^1 g(x,Q)\,dx$ (the zeroth moment of $g$) is formally infrared
divergent. This divergence was absent in the original analysis
\cite{KharzeevSatz:1994}, which used the simplified parametrization
$g(x)=C_g(\nu+1)(1-x)^\nu$, where the large-$x$ exponent $\nu$ was denoted
by $k$ in that reference. For this form, $A_2$ is finite. In the
present work we therefore restrict the sum-rule analysis to $n\ge 4$, for
which the Mellin integrals converge; the cutoff dependence of the excluded
$n=2$ profile is illustrated in Fig.~\ref{fig:n2_appendix}. Higher moments
$A_{n\ge 4}$ are increasingly sensitive to the large-$x$ tail of the
distribution.

The moments $A_n$ defined above correspond to the leading contraction of the
spin-$n$ gluonic operator with the external quarkonium momentum. For a massive
target, however, the symmetric spin-$n$ tensor contains trace-subtraction terms
proportional to powers of $m_N^2$. Neglecting these terms amounts to treating
the nucleon as effectively massless in the operator matrix element, which is
justified only when the target-mass parameter $\tau=m_N^2/(4\epsilon_0^2)$ is
small. In the original treatment \cite{KharzeevSatz:1994}, these trace terms
were neglected, leading to the sum rule
\begin{equation}
\int_0^1 dy\, y^{n-2}\sqrt{1-y^2}\,
\sigma_{\Phi N}\!\left(s(y)\right)
=
\kappa^{n-1} I(n)\,A_n(Q),
\label{eq:sr_no_tmc}
\end{equation}
where $\kappa\equiv m_N/\epsilon_0$ is the dimensionless ratio of the nucleon
mass to the quarkonium binding-energy scale. It converts the hadronic
dispersion variable $y=m_N/\lambda$ into the partonic variable
$z=\kappa x/y$. Here $m_N=0.938$~GeV is the nucleon mass,
$\lambda=(s-M_\Phi^2-m_N^2)/(2M_\Phi)$ is the
nucleon energy in the quarkonium rest frame ($M_\Phi=3.097$~GeV for $J/\psi$),
$s(y)=M_\Phi^2+m_N^2+2M_\Phi m_N/y$
is the centre-of-mass energy squared, the factor $\sqrt{1-y^2}=|\vec{p}_N|/\lambda$
is the three-momentum of the nucleon divided by its energy $\lambda$ in the
quarkonium rest frame arising in the dispersion-relation
derivation \cite{KharzeevSatz:1994}, and
\begin{equation}
I(n)=
2\pi^{3/2}\left(\frac{16}{3}\right)^2
\frac{\Gamma\!\left(n+\frac52\right)}{\Gamma(n+5)}
\left(\frac{4}{3\alpha_s}\right)\frac{1}{m_Q^2}.
\end{equation}
The factor $\kappa^{n-1}$ in Eq.~\eqref{eq:sr_no_tmc} follows from this
change of variables and is common to the no-TMC and TMC moment normalizations. Equation~\eqref{eq:sr_no_tmc} is quoted
using the convention of \cite{KharzeevSatz:1994} as the historical no-trace
sum rule used in that work. The on-shell-domain no-TMC density employed in the
comparisons below is constructed separately in Eq.~\eqref{eq:physical_density_notmc}.
The upper endpoint $y=1$ is the external two-body $\Phi N$ kinematic threshold, where
$\lambda=m_N$ and the relative three-momentum vanishes.  Values $y>1$ would
imply $\lambda<m_N$ and hence an imaginary relative momentum; they belong only
to the subthreshold analytic continuation and not to the on-shell incoming
domain. This kinematic statement does not by itself locate the opening of a
complete hadronic inelastic channel.

For $J/\psi$, the numerical value
$\tau=m_N^2/(4\epsilon_0^2)\simeq 8.6$ is large. Already at $x=0.5$,
$\tau x^2\simeq2.1$, so the trace terms are not perturbatively small in the
large-$x$ region. The traceless tensor structure generates a double series in
operator spin and in the number of trace contractions. Earlier treatments often
reorganised this series into a factorised fixed-spin moment. That representation
is useful as an algebraic check, but the numerical predictions below do not rely
on it: the physical restriction $y\leq1$ couples the partonic and spectral
variables and must be imposed after the complete trace series has been resummed.
We therefore work directly with the all-trace-order generating function and its physical
discontinuity.

\subsection{From the Wilson series to the exact all-trace-order kernel}
\label{sec:allj_derivation}

The OPE in Eq.~\eqref{eq:OPE_amplitude} is naturally written as a discrete
series in the even operator spin $n$.  Physical cross sections, however, are
obtained from the discontinuity of the fully resummed forward amplitude, not
from any finite subset of polynomial terms.  It is therefore useful to first
reconstruct the known no-TMC amplitude in a form whose analytic structure is
explicit, and only then to include the complete tower of target-mass trace
contractions.  This also provides a direct normalization and branch-cut check
for the TMC result.  The no-TMC resummation follows \cite{Kharzeev:1999};
the all-trace-order resummation and its physical continuation are derived
explicitly below rather than assumed from that reference.

We collect the Coulombic Wilson coefficients into the no-TMC generating
function
\begin{equation}
 F_0(z)=\sum_{n=2,4,\ldots}d_n z^{n-2},
 \label{eq:F0_definition}
\end{equation}
where, in the physical convolution, $z=x\lambda/\epsilon_0$.  The point
$z=1$ is therefore the partonic dissociation boundary $x\lambda=\epsilon_0$.
Using the explicit coefficients $d_n$, the series can be summed in closed form,
\begin{equation}
 F_0(z)=\frac{C\pi}{z^6}\left[\frac23\bigl((1-z)^{3/2}+(1+z)^{3/2}\bigr)-\frac43-\frac{z^2}{2}-\frac{z^4}{32}\right],
 \qquad C=\left(\frac{32}{N_c}\right)^2.
 \label{eq:F0_exact}
\end{equation}
The branch cut beginning at $z=1$ is what generates the absorptive part.  Its
physical discontinuity is
\begin{equation}
 \phi_0(z)=\frac{1}{2i}\Disc F_0(z)
 =\frac{2C\pi}{3}\frac{(z-1)^{3/2}}{z^6}\,\theta(z-1).
 \label{eq:phi0_exact}
\end{equation}
Equation~\eqref{eq:phi0_exact} reproduces the standard no-TMC partonic kernel
and will serve as the building block for the exact trace-resummed result.

For a massive target, every spin-$n$ matrix element contains a finite set of
trace contractions.  Denoting their number by $j$ and introducing
$b=\tau x^2$, the complete amplitude may be organised as
\begin{equation}
 \mathcal F_{\rm TMC}(z,b)
 =\sum_{j=0}^{\infty}\left(-\frac{b}{z^2}\right)^j S_j(z),
 \qquad
 S_j(z)=\sum_{\substack{m\ge 2+2j\\ m\,\mathrm{even}}}
 \binom{m-j}{j}d_m z^{m-2}.
 \label{eq:double_trace_series}
\end{equation}
This representation makes clear why a term-by-term finite-spin prescription is
insufficient: individual finite-spin terms are analytic and do not possess an
independent physical cut. The branch cut, and hence the absorptive cross
section, appears only after the complete spin series has been resummed. The
infrared sensitivity of $A_2$ therefore does not imply a separately identifiable
or dominant $n=2$ contribution to the finite-energy cross section. At finite
energy the positive lower limits in Eqs.~\eqref{eq:xmin_baseline} and
\eqref{eq:xmin_shifted} render the resummed convolutions finite. The
$n=4,6,8,\ldots$ moments used below diagnose weighted integral properties of
the same resummed kernel; they are not a truncation of $\sigma_{\Phi N}$.
The fixed-$j$ decomposition in Sec.~\ref{sec:fixedj_spread} instead tests the
convergence of the kinematic trace expansion at fixed twist.

The spin and trace sums must be interchanged and resummed before analytic
continuation.  Performing the finite combinatorial sum at fixed $m$ gives
\begin{equation}
 \mathcal F_{\rm TMC}(z,b)=
 \frac{r_+^3F_0(zr_+)-r_-^3F_0(zr_-)}{v}
 +\frac{b}{z^2}F_0(\sqrt{-b}),
 \label{eq:Ftmc_exact}
\end{equation}
where
\begin{equation}
 v=\sqrt{1-\frac{4b}{z^2}},\qquad
 r_\pm=\frac{1\pm v}{2}.
 \label{eq:rpm_exact}
\end{equation}
The second term in Eq.~\eqref{eq:Ftmc_exact} removes the endpoint that would
otherwise extend the finite trace sum by one spurious contribution; away from
$z=0$ it has no discontinuity on the physical cut.  The closed form connects
smoothly to the known theory: it reduces to $F_0(z)$ when $b\to0$, and its
small-$b$ expansion reproduces the independently derived fixed-trace-order
functions and their spectral kernels.  These relations are displayed explicitly
in \ref{app:fixed_j_expansion}.

In physical kinematics, $z=x\lambda/\epsilon_0$ and $b=\tau x^2$, so the
combination entering the square root is independent of $x$,
\begin{equation}
 \frac{4b}{z^2}=\frac{m_N^2}{\lambda^2}=y^2,
 \qquad
 v=\sqrt{1-y^2}.
 \label{eq:physical_delta}
\end{equation}
This identity is the key simplification that allows the resummed amplitude to
be converted into a closed physical convolution.

\subsection{Analytic continuation, shifted spectral baseline, and exact moment}
\label{sec:physical_kernel}

The next step is to continue the resummed amplitude through its physical cut.
The two arguments of the no-TMC function are
$u_\pm=zr_\pm$.  Along the real physical cut they move in opposite directions:
$du_+/dz>0$, whereas $du_-/dz<0$.  Consequently, the reversed orientation of
the $u_-$ cut changes the apparent relative minus sign in
Eq.~\eqref{eq:Ftmc_exact} into a plus in the spectral density.  The exact
all-orders physical kernel is therefore
\begin{equation}
 \Phi_{\rm TMC}(z,b)=
 \frac{r_+^3\phi_0(zr_+)+r_-^3\phi_0(zr_-)}{v}.
 \label{eq:Phi_exact}
\end{equation}
The two branches therefore contribute coherently to the absorptive amplitude.
In particular, the trace corrections must be resummed before the discontinuity
is taken; applying a fixed-spin weight to an already continued no-TMC kernel
would miss this branch structure.

Adopting the conventional Coulombic normalization used in
\cite{Kharzeev:1996yx,Kharzeev:1999},
\begin{equation}
 K_\sigma=\frac{4}{3\alpha_s m_Q^2},
\end{equation}
the exact TMC spectral baseline can be written as a sum of two shifted no-TMC
convolutions. In the strict Coulombic limit this normalization is equivalent to
$K_\sigma=a_0^3\epsilon_0$. In the phenomenological treatment adopted below,
$\epsilon_0$ is varied independently while $m_Q$, $\alpha_s$, and hence
$K_\sigma$, are held fixed. The convention has no effect on the reported
TMC/no-TMC ratios or normalized shapes, in which the common factor cancels:
\begin{equation}
 \sigma_{\rm TMC}(\lambda)=
 \frac{r_+^2\sigma_0(\lambda r_+)+r_-^2\sigma_0(\lambda r_-)}{\sqrt{1-y^2}},
 \qquad y=\frac{m_N}{\lambda},
 \label{eq:sigma_shift}
\end{equation}
where
\begin{equation}
 \sigma_0(\lambda)=K_\sigma
 \int_{\epsilon_0/\lambda}^{1}\frac{dx}{x}\,g(x,Q)\,z\phi_0(z),
 \qquad z=\frac{x\lambda}{\epsilon_0}.
 \label{eq:sigma0_exact}
\end{equation}
The convolution is defined to vanish when its energy argument $E$ satisfies
$E\leq\epsilon_0$; equivalently, every occurrence of $\sigma_0(E)$ includes the
support factor $\Theta(E-\epsilon_0)$. Thus the TMC result is not a truncated
modification of the known convolution: it is the discontinuity of the fully
resummed trace series and requires no spin-order or trace-order cutoff.

Equation~\eqref{eq:sigma_shift} also fixes the strict two-body-threshold
behaviour. With $v\equiv\sqrt{1-y^2}=|\vec p_N|/\lambda$, one has
$r_\pm\to1/2$ as $v\to0$, and therefore
\begin{equation}
 \sigma_{\rm TMC}(\lambda)\underset{v\to0^+}{\sim}
 \frac{\sigma_0(m_N/2)}{2v}.
 \label{eq:threshold_inverse_velocity}
\end{equation}
For the central Coulombic parameters $m_N/2>\epsilon_0$, so within the
adopted Coulombic OPE the partonic absorptive cut is already open at zero
incoming $J/\psi N$ relative momentum. The resulting $1/v$ behaviour has the form of a
flux enhancement and is integrable in the sum rule because the physical moment
contains the compensating factor $v\,\sigma_{\rm TMC}$; the finite quantity
entering the dispersion moment is therefore $v\sigma_{\rm TMC}$. The point
$y=1$ is, however, the incoming two-body endpoint rather than a model-independent
hadronic inelastic threshold. The inverse-velocity limit and any non-monotonic
structure in the unresolved endpoint layer should therefore be interpreted as
properties of the leading-twist partonic/OPE spectral baseline, not as a
complete hadronic-threshold prediction. The numerical curves begin at
$\sqrt{s}=4.045$~GeV, $10$~MeV above the endpoint, and display only the resolved
behaviour outside the asymptotic $v\to0$ layer.

For the moment analysis it is useful to derive a second, independently
integrable representation.  Reordering the physical $x$, $y$, and spectral
integrations yields
\begin{align}
 M_{n,\mathrm{phys}}^{\rm TMC}
 &=K_\sigma\kappa^{n-1}\int_0^1dx\,x^{n-2}g(x,Q)
 \int_1^\infty dz\,z^{-(n-1)}\phi_0(z)
 \frac{|1-q^2|}{(1+q^2)^{n+2}},\label{eq:physical_weight}\\
 q&=\frac{\kappa x}{2z},\qquad \kappa=\frac{m_N}{\epsilon_0}.
\end{align}
Here and below, ``physical moment'' means that the dispersion integral is
restricted to on-shell external kinematics $0<y\leq1$; it does not mean that
the moment itself is directly measurable or that a complete hadronic inelastic
channel necessarily opens at $y=1$. The factor multiplying $\phi_0$ is
the exact physical-domain TMC weight after all trace orders have been resummed.  Unlike a factorised fixed-spin weight, it
retains the correlation between the partonic variable and the physical
$y\leq1$ boundary.  The same moment may also be obtained either from the direct $y$ moment of
Eq.~\eqref{eq:sigma_shift} or from the spectral kernel
Eq.~\eqref{eq:Phi_exact}.  Their agreement provides an independent closure of
the physical-domain construction; numerical details are summarised in
Sec.~\ref{sec:sensitivities}.

\subsection{Relation to the fixed-spin hypergeometric sum rule}
\label{sec:hypergeom_relation}
The hypergeometric weight derived in \cite{Kharzeev:1996yx} is
algebraically correct at fixed operator spin.  The present analysis shows that
this factorised form corresponds to the analytically extended dispersion
moment, before the physical restriction $y\leq1$ is imposed.  Summing the
trace contractions before the spin sum gives
\begin{equation}
 \widetilde A_n^{\rm ext}(Q)=\int_0^1 dx\,x^{n-2}g(x,Q)\,T_n(x),
\end{equation}
with
\begin{equation}
 T_n(x)={}_3F_2\!\left(\frac54+\frac n2,\frac74+\frac n2,1+n;
 \frac{5+n}{2},3+\frac n2;-\tau x^2\right).
\label{eq:hypergeom_weight}
\end{equation}
The superscript ``ext'' is essential. This factorised moment is obtained when
the fixed-$n$ trace series is converted into a dispersion moment before the
physical restriction $y\leq1$ is imposed.  Equivalently, it is the exact
all-trace-order result on the analytically extended dispersion domain.  On the
physical domain, however, the boundary becomes
$z_{\min}(x)=\max(1,\kappa x)$ and couples the partonic and spectral
variables.  The $x$ integral therefore no longer factorises into a universal
weight $T_n(x)$ multiplying the bare Mellin moment; the exact physical result
is instead Eq.~\eqref{eq:physical_weight}.

This distinction does not invalidate the tensor algebra or the formal sum rule
in \cite{Kharzeev:1996yx}.  It identifies the domain on which that
hypergeometric representation is exact and shows why it cannot by itself be
used as the physical near-threshold moment.  The no-TMC construction in
\cite{Kharzeev:1999} is also unaffected: its Wilson series is resummed exactly and yields the
standard direct convolution.  The new element here is the simultaneous
all-spin, all-trace-order resummation followed by the physical discontinuity, which
retains the $y\leq1$ boundary. The derivational bridge between the extended-domain kernel moments and Eq.~\eqref{eq:hypergeom_weight} is given in~\ref{app:extended_hypergeom}.

\subsection{PDF inputs, uncertainties, and $x$-resolved diagnostics}
All central and uncertainty results use native LHAPDF grids \cite{LHAPDF6} for
ABMP16, MSHT20, CT18 and NNPDF4.0, with direct numerical integration of the
tabulated gluon distribution. The exact set identifiers are listed in
Table~\ref{tab:lhapdf_sets}.

\begin{table}[H]
\centering
\caption{LHAPDF inputs used in the numerical analysis. All sets are NNLO.}
\label{tab:lhapdf_sets}
\small
\begin{tabular}{llll}
\toprule
PDF family & LHAPDF identifier & $N_{\rm members}$ & Uncertainty prescription \\
\midrule
ABMP16  & \texttt{ABMP16\_5\_nnlo}          & 30  & Symmetric Hessian, 29 error members \\
MSHT20  & \texttt{MSHT20nnlo\_as118}        & 65  & Hessian, 32 eigenvector pairs \\
CT18    & \texttt{CT18NNLO}                 & 59  & Hessian, 29 eigenvector pairs; 90\% to 68\% CL \\
NNPDF4.0 & \texttt{NNPDF40\_nnlo\_as\_01180} & 101 & Monte Carlo replicas \\
\bottomrule
\end{tabular}
\end{table}

The identifier \texttt{NNPDF40\_nnlo\_as\_01180} denotes the standard
NNPDF4.0 fitted-charm ensemble \cite{NNPDF40main}; it is not the separate perturbative-charm
variant.  Only the gluon distribution enters the present calculation.

For cross-checking and visualisation purposes we retain analytic five-parameter
approximations of the form
\begin{equation}
xg(x,Q)=A\,x^{\alpha}(1-x)^{\beta}
\bigl(1+\epsilon \sqrt{x}+\gamma x\bigr),
\label{eq:pdf_param}
\end{equation}
whose parameters are listed in \ref{app:pdf_params}. They are not used for
the tabulated physical moments, cross sections, uncertainty bands, parameter
scans, or any quoted endpoint location.

PDF uncertainties are propagated using:
\begin{itemize}
\item ABMP16 symmetric Hessian: 29 error members,
$\delta X=[\sum_{k=1}^{29}(X^k-X^0)^2]^{1/2}$.
\item MSHT20 Hessian: 32 eigenvector pairs,
$\delta X=\tfrac12[\sum_{k=1}^{32}(X^{+k}-X^{-k})^2]^{1/2}$.
\item CT18 Hessian: the same paired formula, divided by $1.645$ to convert the
native 90\% confidence level to 68\% confidence level.
\item NNPDF4.0 Monte Carlo: standard deviation over 100 replicas.
\end{itemize}
The Coulombic OPE coefficients retain the fixed value $\alpha_s=0.3$; LHAPDF
is used only for the evolved gluon distributions. Uncertainties are propagated
from the complete ensembles of all four PDF families. For visual clarity,
Figs.~\ref{fig:sigma_exact} and~\ref{fig:sigma_tmc_ratio} show the CT18 band,
while Fig.~\ref{fig:sigma_tmc_ratio} also displays the envelope of the four
central predictions; the omission of the other three set-specific bands from
these figures is graphical only. No PDF band is shown in
Fig.~\ref{fig:photoproduction-context}, because that figure is included solely
as a normalized shape comparison and is not used for a quantitative fit.

At this stage the Mellin moments are treated as diagnostic objects in their own
right. To make the PDF support explicit, define
\begin{equation}
w_n(x;Q)=x^{n-2}g(x,Q),\qquad
\rho_n(x;Q)=\frac{x\,w_n(x;Q)}{A_n(Q)}
=\frac{x^{n-2}xg(x,Q)}{A_n(Q)}.
\label{eq:rho_n}
\end{equation}
The log-$x$ density $\rho_n$ integrates to unity with respect to $d\ln x$ and
shows how the unweighted Mellin support shifts from small $x$ at low $n$ toward
intermediate and large $x$ at higher $n$.

\begin{figure}[H]
\centering
\includegraphics[width=0.98\textwidth]{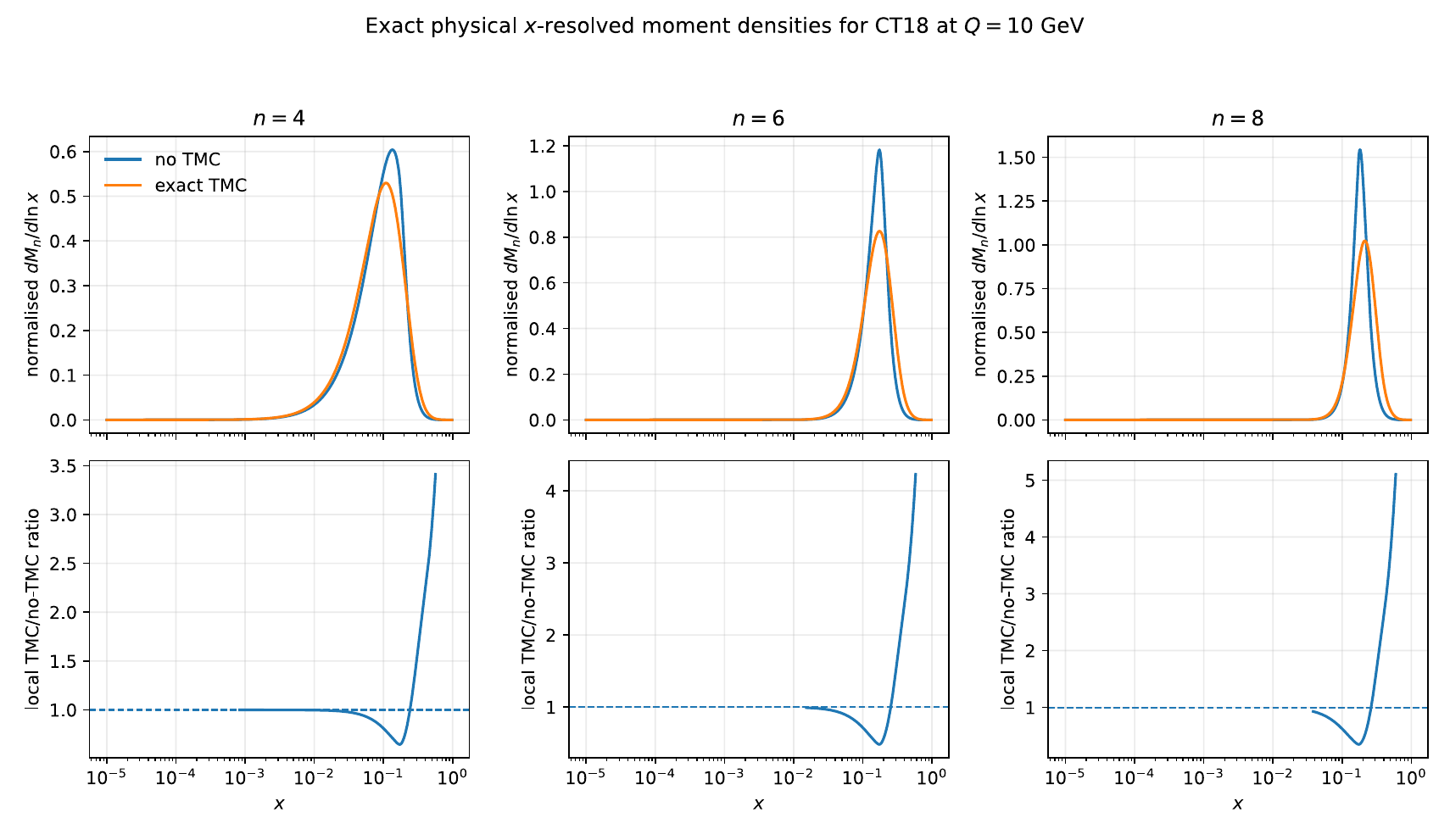}
\caption{Exact physical $x$-resolved moment densities for the CT18 central grid
at $Q=10$~GeV.  The upper panels compare $\rho_{n,\rm phys}^{(0)}$ and
$\rho_{n,\rm phys}^{\rm TMC}$, each normalised to unit area, for $n=4,6,8$.  The lower panels
show the local ratio of the unnormalised TMC and no-TMC densities; points for
which the no-TMC density is below $10^{-3}$ of its maximum are masked to avoid
a ratio-of-vanishing-functions endpoint artefact.  The figure therefore
separates a reshaping of the sampled $x$ region from the net moment suppression.}
\label{fig:xresolved_physical}
\end{figure}

The exact physical TMC moment also admits an $x$-resolved density,
\begin{equation}
 w_{n,\rm phys}^{\rm TMC}(x;Q)=K_\sigma\kappa^{n-1}x^{n-2}g(x,Q)
 \int_1^\infty dz\,z^{-(n-1)}\phi_0(z)
 \frac{|1-q^2|}{(1+q^2)^{n+2}},
\label{eq:physical_density}
\end{equation}
with $q=\kappa x/(2z)$.  The corresponding physical no-TMC density is
not obtained simply by setting the last factor to unity.  Before the trace
resummation, the condition $y\leq1$ appears as the $x$-dependent lower boundary
\begin{equation}
 w_{n,\rm phys}^{(0)}(x;Q)=K_\sigma\kappa^{n-1}x^{n-2}g(x,Q)
 \int_{\max(1,\kappa x)}^\infty dz\,z^{-(n-1)}\phi_0(z)
 \sqrt{1-\left(\frac{\kappa x}{z}\right)^2}.
\label{eq:physical_density_notmc}
\end{equation}
The square-root factor in Eq.~\eqref{eq:physical_density_notmc} is the
physical phase-space factor $\sqrt{1-y^2}$ after using $y=\kappa x/z$.
The normalized densities per $d\ln x$ plotted below are
\begin{equation}
 \rho_{n,\rm phys}^{\rm TMC}(x;Q)=
 \frac{x\,w_{n,\rm phys}^{\rm TMC}(x;Q)}{M_{n,\rm phys}^{\rm TMC}},
 \qquad
 \rho_{n,\rm phys}^{(0)}(x;Q)=
 \frac{x\,w_{n,\rm phys}^{(0)}(x;Q)}{M_{n,\rm phys}^{(0)}},
 \label{eq:rho_physical}
\end{equation}
where $M_{n,\rm phys}^{a}=\int_0^1dx\,w_{n,\rm phys}^{a}(x;Q)$, so that
$\int d\ln x\,\rho_{n,\rm phys}^{a}=1$ for $a=(0),{\rm TMC}$.
The exact TMC mapping converts this moving boundary and phase-space factor into
the correlated weight in Eq.~\eqref{eq:physical_density}, with a fixed lower
limit $z=1$.  Their
local ratio therefore compares two physical densities whose spectral support
has been reorganised, rather than two expressions differing only by a
multiplicative factor.  Figure~\ref{fig:xresolved_physical} displays both quantities for representative CT18 moments. The unnormalised local ratio shows increasing suppression over the dominant support as $n$ grows, while the separately normalised upper panels display the accompanying redistribution in $x$. A compensating enhancement appears in the extreme large-$x$ tail where the absolute density is already negligible. This gives the local $x$-space origin of the integrated hierarchy found below.

\begin{figure}[p]
\centering
\includegraphics[width=0.98\textwidth]{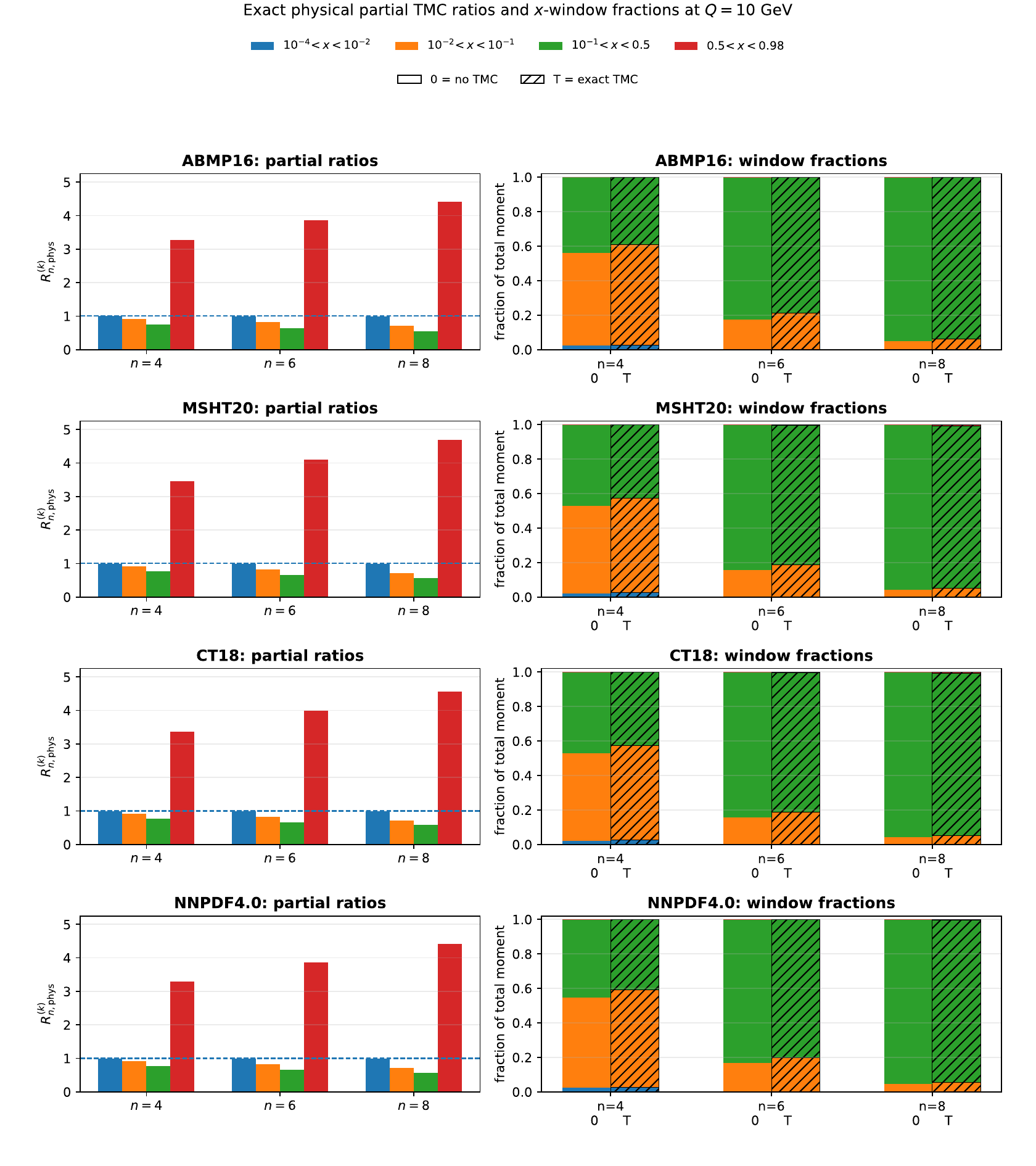}
\caption{Exact physical partial ratios and their fractional contribution to the corresponding total moment for four
$x$ windows, the four central PDF grids, and $n=4,6,8$ at $Q=10$~GeV.  The
left-hand panels show $R_{n,\rm phys}^{(k)}$.  The right-hand panels show the
same windows as fractions of the corresponding total moment; for each $n$, the
left bar ($0$) is the no-TMC decomposition and the right hatched bar ($T$) is
the exact-TMC decomposition ($0=$ no TMC, $T=$ exact TMC).  Each stacked bar is
normalised independently to its corresponding total moment.  Although the ratio in the largest-$x$ interval
exceeds unity, its fraction of the total moment is extremely small.  The net
suppression is generated in the small-to-intermediate-$x$ region, which carries essentially
all of the physical weight.  The two smallest-$x$ fractions become visually negligible for the higher moments.}
\label{fig:partial_physical_ratios}
\end{figure}

To display the same information in the finite windows used in the earlier
analysis, define
\begin{equation}
 R_{n,\rm phys}^{(k)}=
 \frac{\displaystyle\int_{x_k^-}^{x_k^+}dx\,w_{n,\rm phys}^{\rm TMC}(x;Q)}
 {\displaystyle\int_{x_k^-}^{x_k^+}dx\,w_{n,\rm phys}^{(0)}(x;Q)}.
\label{eq:partial_physical_ratio}
\end{equation}
Figure~\ref{fig:partial_physical_ratios} is the exact physical-domain counterpart
of the fixed-spin partial-ratio diagnostic.  The small-$x$ windows remain close to
unity, while the small-to-intermediate-$x$ region supplies the net suppression.
The largest-$x$ window is instead enhanced.  This does not imply an enhancement of
the full moment: that window carries little no-TMC weight, and the physical
all-trace-order mapping redistributes spectral support across the $x$-dependent
boundary.  The integrated result remains below unity, as shown below.

The partial ratio exceeds unity in the largest-$x$ interval because the
corresponding no-TMC contribution is already extremely small owing to the
$x$-dependent physical boundary $z_{\min}=\max(1,\kappa x)$ and the falling
gluon PDF.  The exact all-trace-order kernel redistributes a small amount of spectral
strength into this endpoint region.  This enhancement is negligible in absolute
magnitude and does not compensate the suppression generated in the dominant
small-to-intermediate-$x$ region.  To make this distinction visible, the right-hand
panels of Fig.~\ref{fig:partial_physical_ratios} show each window as a fraction
of the corresponding total no-TMC or exact-TMC moment.  Each stacked bar is
normalised to its own total moment; the overall suppression
$M_{n,\rm phys}^{\rm TMC}/M_{n,\rm phys}^{(0)}<1$ is shown separately in
Fig.~\ref{fig:physical_moment_ratios}.

\section{Direct convolution and numerical results}
\label{sec:results}
\subsection{Exact all-trace-order direct convolution and threshold structure}
\label{sec:threshold_structure}

Equation~\eqref{eq:sigma_shift} turns the resummed trace series into a direct
finite-energy leading-twist spectral baseline.  The unshifted no-TMC baseline convolution
Eq.~\eqref{eq:sigma0_exact} has the lower integration limit
\begin{equation}
 x_{\min}^{(0)}=\frac{\epsilon_0}{\lambda}.
 \label{eq:xmin_baseline}
\end{equation}
In the exact TMC expression, the two shifted convolutions instead have the
branch-specific limits
\begin{equation}
 x_{\min,\pm}=\frac{\epsilon_0}{\lambda r_\pm}.
 \label{eq:xmin_shifted}
\end{equation}
These limits select the gluon momentum fractions that can satisfy the partonic
dissociation condition in each branch. Consequently every convolution remains
finite at finite energy, even though the formal Mellin moment
$A_2=\int_0^1 g(x,Q)\,dx$ is infrared sensitive for modern small-$x$ gluons.
The two statements refer to different limits: $A_2$ probes the distribution all
the way to $x=0$, whereas the finite-energy cross section always retains
positive kinematic cutoffs.  The convergent moments with $n\ge4$ therefore
provide clean integral diagnostics of the same all-spin kernel without being
interpreted as a truncation of the cross section.

Two kinematic boundaries enter the leading-twist construction with distinct
roles. The external incoming two-body endpoint is
\begin{equation}
 \sqrt{s_{\rm ext}}=M_{J/\psi}+m_N,
 \qquad \lambda_{\rm ext}=m_N,
\end{equation}
and is the endpoint of the physical incoming $J/\psi N$ domain. The numerical curves shown below begin at $\sqrt{s}=4.045$~GeV, $10$~MeV
above the endpoint, and do not resolve the asymptotic endpoint layer. Inside the convolution, the condition
\begin{equation}
 x\lambda\geq\epsilon_0
\end{equation}
sets the partonic support.  Since $\epsilon_0<m_N$ for the central parameters,
partonic support within the Coulombic OPE convolution is already open at the
external threshold; increasing the energy lowers the unshifted baseline limit
$x_{\min}^{(0)}$ and the two shifted limits $x_{\min,\pm}$, moving the
convolutions toward the small-$x$ region where target-mass effects become weak.
Here $M_{J/\psi}+m_N$ is the incoming $J/\psi N$ two-body kinematic endpoint;
it should not be identified with a complete hadronic open-charm or multihadron
dissociation threshold. The quantity $\sigma_{\Phi N}$ generated by the
leading-twist convolution is the discontinuity of the forward $J/\psi N$
amplitude associated with partonic dissociation in the Coulombic OPE. In the
finite-energy range considered below, it is used as a leading-twist reference
for the energy dependence of this absorptive contribution. Arbitrarily close
to the incoming endpoint, however, the calculation does not include the actual
hadronic-channel thresholds, higher-twist multipole terms, finite-size effects,
or final-state dynamics; it therefore cannot be identified with the complete
physical inelastic $J/\psi N$ cross section in that layer. It also does not
describe elastic $J/\psi N$ scattering, which is a different channel and is not
determined by the absorptive discontinuity alone. Nor is it the exclusive
$\gamma p\to J/\psi p$ cross section, whose construction additionally requires
the photon--$J/\psi$ conversion, the real part of the amplitude, and the
momentum-transfer dependence.

\subsection{Moment and spectral-baseline results}
\begin{figure}[H]
\centering
\includegraphics[width=0.78\textwidth]{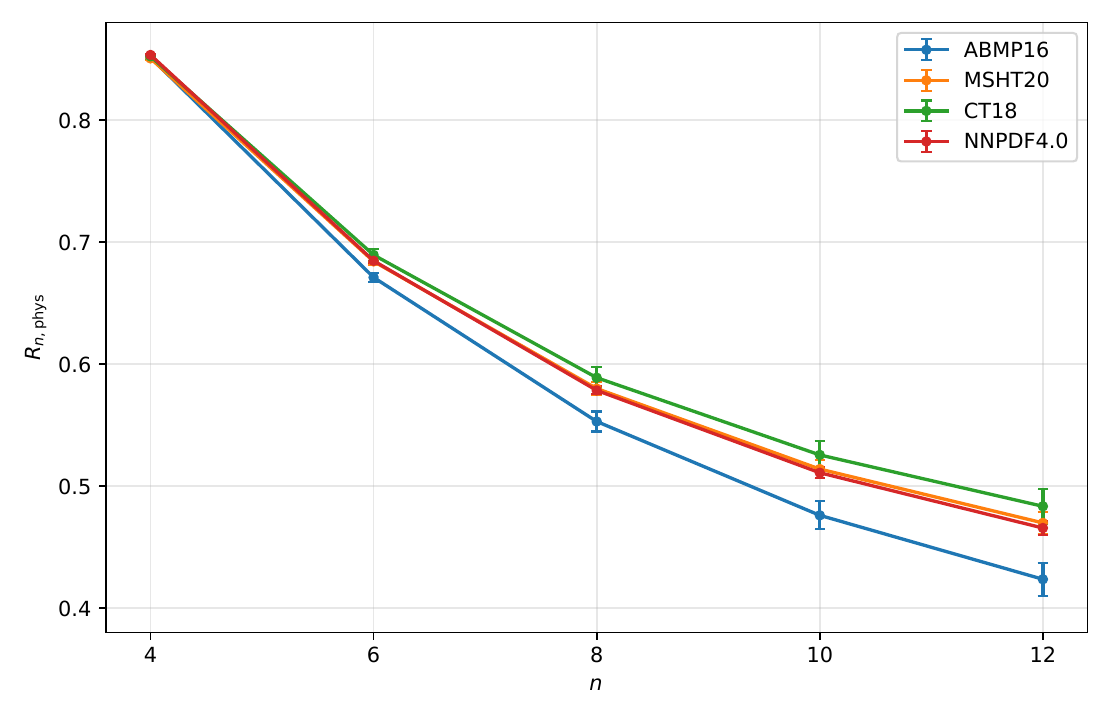}
\caption{Exact on-shell-domain TMC/no-TMC moment ratios for the four global-fit
PDF families.  Error bars show the propagated 68\% CL-equivalent PDF
uncertainties.}
\label{fig:physical_moment_ratios}
\end{figure}

Before examining the finite-energy spectral baseline, we first quantify the
integrated effect of the target-mass correction through the on-shell-domain
moments. These moments are not direct observables; rather, they provide a
controlled diagnostic of how the trace resummation modifies the gluon-PDF
contribution as the operator spin $n$ increases. Since higher moments place
progressively greater weight on the large-$x$ region, their ratios reveal both
the systematic $n$-dependence of the correction and its sensitivity to the
different global PDF families. We therefore define
\begin{equation}
 R_{n,\mathrm{phys}}=\frac{M_{n,\mathrm{phys}}^{\rm TMC}}{M_{n,\mathrm{phys}}^{(0)}}.
\end{equation}
At $Q=10$~GeV and $\epsilon_0=0.16$~GeV, all four PDF families give
$R_{4,\mathrm{phys}}\simeq0.85$.  This near universality reflects the broad,
moderate-$x$ support of the lowest convergent moment.  As $n$ increases, the
moment shifts toward larger $x$, the target-mass reorganisation becomes stronger,
and the PDF dependence becomes more visible: by $n=12$ the ratio lies between
about $0.42$ and $0.48$.  The complete ratio pattern is shown in
Fig.~\ref{fig:physical_moment_ratios} and Table~\ref{tab:physical_ratios}.

\begin{table}[H]
\centering
\caption{Exact on-shell-domain TMC/no-TMC moment ratios at $Q=10$~GeV and $\epsilon_0=0.16$~GeV. Uncertainties are 68\% CL equivalents.}
\label{tab:physical_ratios}
\resizebox{\linewidth}{!}{%
\begin{tabular}{lrrrrr}
\toprule
Set & $R_4$ & $R_6$ & $R_8$ & $R_{10}$ & $R_{12}$\\
\midrule
ABMP16 & $0.8514\pm0.0008$ & $0.6711\pm0.0038$ & $0.5530\pm0.0082$ & $0.4761\pm0.0114$ & $0.4237\pm0.0136$ \\
MSHT20 & $0.8506\pm0.0010$ & $0.6842\pm0.0029$ & $0.5800\pm0.0054$ & $0.5140\pm0.0075$ & $0.4699\pm0.0091$ \\
CT18   & $0.8523\pm0.0018$ & $0.6895\pm0.0046$ & $0.5889\pm0.0085$ & $0.5256\pm0.0116$ & $0.4835\pm0.0141$ \\
NNPDF4.0 & $0.8534\pm0.0008$ & $0.6847\pm0.0020$ & $0.5785\pm0.0033$ & $0.5110\pm0.0045$ & $0.4657\pm0.0055$ \\
\bottomrule
\end{tabular}}
\end{table}

\begin{figure}[H]
\centering
\includegraphics[width=0.82\textwidth]{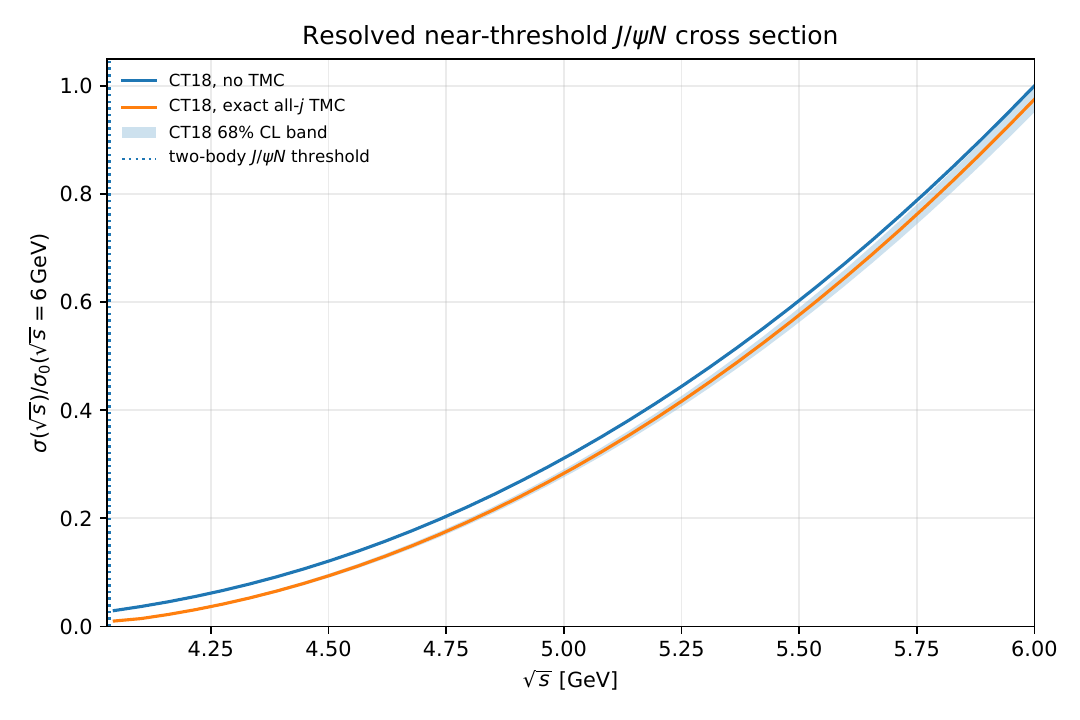}
\caption{Resolved leading-twist all-trace-order $J/\psi N$ spectral baseline for CT18 at
$Q=10$~GeV, with and without TMC.  Both curves are divided by $\sigma_0(\sqrt{s}=6~\mathrm{GeV})$; the band is the CT18 68\% CL-equivalent Hessian
uncertainty on the TMC result. The displayed interval extends from the first
resolved point at $\sqrt{s}=4.045$~GeV to $6$~GeV. The strict endpoint limit is
described by Eq.~\eqref{eq:threshold_inverse_velocity} and is not interpreted as
a complete hadronic-threshold prediction.}
\label{fig:sigma_exact}
\end{figure}

\begin{figure}[H]
\centering
\includegraphics[width=0.82\textwidth]{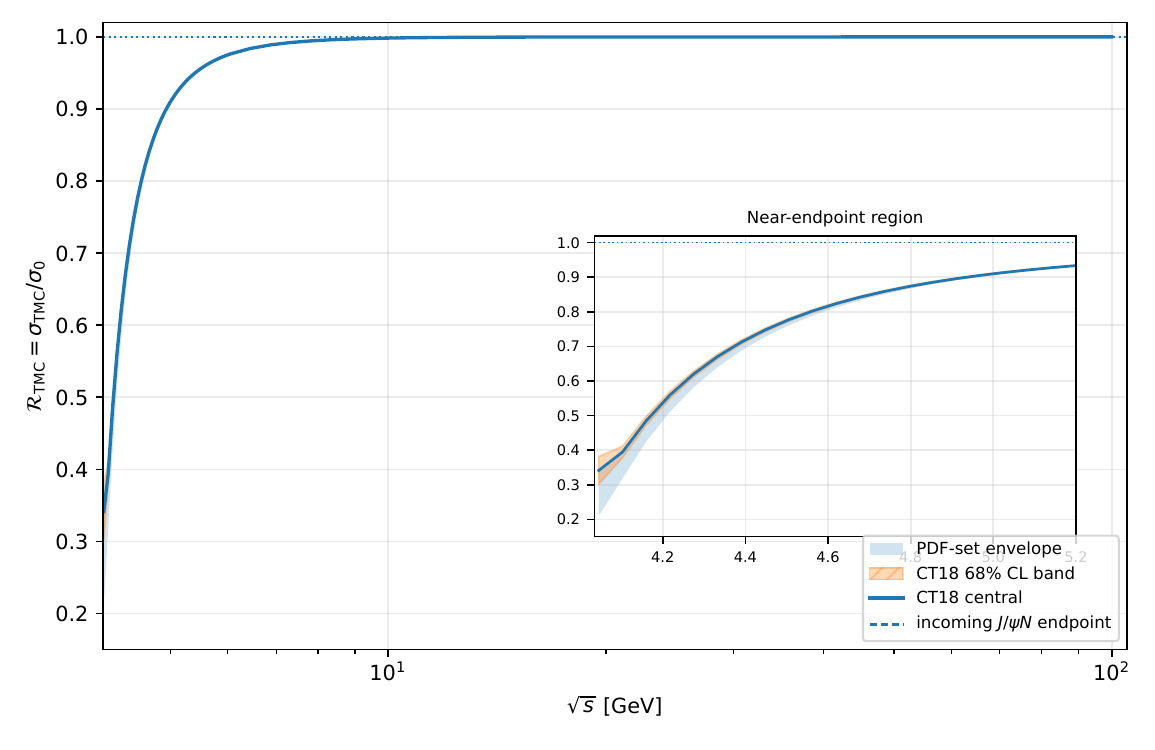}
\caption{Exact all-trace-order ratio $\mathcal{R}_{\mathrm{TMC}}=\sigma_{\mathrm{TMC}}/\sigma_0$.
The light filled band is the envelope of the four central PDF sets, while the hatched band
shows the CT18 68\% CL-equivalent Hessian uncertainty. The inset enlarges the near-endpoint
region where the two bands can be resolved. The numerical curve begins at $\sqrt{s}=4.045$~GeV, about $10$~MeV above
the incoming two-body endpoint, and does not display the asymptotic $1/v$ layer.}
\label{fig:sigma_tmc_ratio}
\end{figure}

The cross-section energy dependence is sharper than the moment hierarchy suggests.
For
\begin{equation}
 \mathcal R_{\rm TMC}(\sqrt{s})=
 \frac{\sigma_{\rm TMC}(\sqrt{s})}{\sigma_0(\sqrt{s})},
\end{equation}
over the tabulated interval above the external incoming two-body endpoint, the strongest
suppression occurs at the lowest energies. At the first tabulated energy,
$\sqrt{s}=4.045$~GeV, the central predictions span roughly $0.21$--$0.34$;
by $\sqrt{s}=4.1025$~GeV they span about $0.32$--$0.39$. This spread reflects
the different large-$x$ gluons sampled in the immediate endpoint region. As
the energy rises, the baseline and shifted lower limits rapidly decrease: the ratio reaches about
$0.91$ near $5$~GeV and is already close to unity by $10$~GeV.  Thus the exact
TMC effect is potentially large in the first resolved partonic/OPE interval above the endpoint,
but it is not a broad renormalisation of the high-energy convolution.  The CT18 cross
sections and the four-set envelope of the ratio are shown in
Figs.~\ref{fig:sigma_exact} and~\ref{fig:sigma_tmc_ratio}.

\subsection{How much of the exact result is carried by the first trace orders?}
\label{sec:fixedj_spread}
The closed all-trace-order expression is the leading-twist spectral baseline, while its fixed-trace
expansion provides a useful way of identifying which part of the near-endpoint
suppression is already visible at low order. The cumulative quantities are defined by
\begin{equation}
 \sigma^{[0:k]}(\lambda)=\sum_{j=0}^{k}\sigma^{(j)}(\lambda),
 \qquad
 \delta_j(\lambda)=\frac{\sigma^{(j)}(\lambda)}{\sigma^{(0)}(\lambda)},
 \label{eq:fixedj_cumulative_main}
\end{equation}
We evaluate the regular terms $j=0,1,2$ with the CT18 central grid at
$Q=10$~GeV and $\epsilon_0=0.16$~GeV.  The explicit kernels are given in \ref{app:fixed_j_expansion}.  Table~\ref{tab:fixedj_spread} shows that
the first trace term produces almost the entire visible correction.  The
$j=2$ contribution is already at the percent level or below except within a
few tens of MeV of the endpoint, and the cumulative $j\leq2$ result follows the
exact resummation to better than one percent from $\sqrt{s}=4.10$~GeV upward.
Only at $4.06$~GeV, where the partonic support is extremely narrow, does the
unresummed higher-$j$ tail still change the result by several percent. The
$4.060$~GeV row is a dedicated fixed-order check below the first energy included
in the main numerical comparison.

\begin{table}[H]
\centering
\caption{Fixed-trace-order content of the CT18 leading-twist spectral baseline at
$Q=10$~GeV and $\epsilon_0=0.16$~GeV.  Here
$\delta_j=\sigma^{(j)}/\sigma^{(0)}$,
$R_{[0:k]}=\sigma^{[0:k]}/\sigma^{(0)}$, and
$\Delta_{[0:2]}=100\,[\sigma^{[0:2]}/\sigma_{\rm exact}-1]$.
The exact all-trace-order result remains the leading-twist baseline; the table does not describe the asymptotic $v\to0$ endpoint layer.}
\label{tab:fixedj_spread}
\footnotesize
\begin{tabular}{rrrrrrr}
\toprule
$\sqrt{s}$ [GeV] & $\delta_1$ & $\delta_2$ & $R_{[0:1]}$ & $R_{[0:2]}$ & $R_{\rm exact}$ & $\Delta_{[0:2]}$ [\%]\\
\midrule
4.060 & $-0.710$ & $+0.0249$ & 0.290 & 0.315 & 0.327 & $-3.65$\\
4.100 & $-0.625$ & $+0.0173$ & 0.375 & 0.392 & 0.390 & $+0.41$\\
4.200 & $-0.467$ & $+0.0069$ & 0.533 & 0.540 & 0.538 & $+0.28$\\
4.400 & $-0.281$ & $+0.0007$ & 0.719 & 0.719 & 0.719 & $+0.06$\\
4.578 & $-0.191$ & $-0.0003$ & 0.809 & 0.809 & 0.808 & $+0.02$\\
5.000 & $-0.0897$ & $-0.0003$ & 0.910 & 0.910 & 0.910 & $+0.003$\\
6.000 & $-0.0249$ & $-0.00004$ & 0.975 & 0.975 & 0.975 & $<0.001$\\
10.000 & $-0.00164$ & $<10^{-6}$ & 0.9984 & 0.9984 & 0.9984 & $<10^{-6}$\\
\bottomrule
\end{tabular}
\end{table}

The pattern is physically simple.  The $j=1$ term lowers the spectral baseline in the resolved near-endpoint
region, while the $j=2$ term provides only a small refinement and changes
sign around $\sqrt{s}\simeq4.5$~GeV.  The rapid convergence away
from the endpoint explains why the exact and no-TMC curves acquire nearly the
same high-energy slope, although their normalized near-endpoint shapes remain
distinct.

\section{Sensitivities and limitations}
\label{sec:sensitivities}
\subsection{Scale, binding-energy, and numerical checks}
The reference PDF-input scale is $Q=10$~GeV, chosen for the present numerical
analysis and for a uniform comparison among the four modern PDF families.
Because the leading-order Coulombic coefficient functions are held fixed, the
scan below measures PDF-input evolution rather than a complete factorisation-scale
uncertainty. To gauge this dependence, the CT18 central member is evaluated at
$Q=2.0,3.1,5.0,10.0,$ and $100$~GeV.  The value $Q=2$~GeV is retained only as a
low-scale boundary check, while $Q=100$~GeV provides an extended test of PDF evolution rather
than a preferred near-threshold charmonium scale.  Throughout this scan,
$\alpha_s=0.3$ and the Coulombic Wilson coefficients are held fixed; only the
evolved gluon distribution is changed. Table~\ref{tab:scale_dependence} shows
that the moment ratios vary moderately, whereas the resolved near-endpoint spectral ratio
is more sensitive because it samples the evolving large-$x$ gluon
distribution.

\begin{table}[h]
\centering
\caption{CT18 PDF-input-scale dependence of the exact physical moment ratios and $\mathcal R_{\rm TMC}(4.1~\mathrm{GeV})$ at fixed leading-order Coulombic coefficients. The $Q=2$~GeV point is a low-scale boundary check, while $Q=100$~GeV provides an extended test of PDF evolution.}
\label{tab:scale_dependence}
\begin{tabular}{ccccc}
\toprule
$Q$ [GeV] & $R_4$ & $R_6$ & $R_8$ & $\mathcal R_{\rm TMC}(4.1)$\\
\midrule
2.0${}^{*}$ & 0.841 & 0.707 & 0.625 & 0.457 \\
3.1 & 0.845 & 0.700 & 0.612 & 0.434 \\
5.0 & 0.848 & 0.695 & 0.601 & 0.413 \\
10.0 & 0.852 & 0.690 & 0.589 & 0.390 \\
100.0 & 0.863 & 0.680 & 0.565 & 0.341 \\
\bottomrule
\end{tabular}
\end{table}

The binding-energy parameter affects both the strength of trace corrections,
through $\tau=m_N^2/(4\epsilon_0^2)$, and the partonic support, through
$\kappa=m_N/\epsilon_0$.  The scan
$\epsilon_0=0.16,0.25,0.40,0.60$~GeV therefore probes more than a simple
rescaling of the kernel. Throughout this scan, $m_Q$, $\alpha_s$, and the
conventional normalization $K_\sigma$ are held fixed; $\epsilon_0$ enters through
$\tau$, $\kappa$, and the partonic support.  As Table~\ref{tab:epsilon0_sensitivity} shows, the
integrated moments change smoothly, while the leading-twist spectral baseline at a fixed energy
close to the endpoint can change dramatically because the accessible partonic
region itself moves.  At fixed $\sqrt{s}=4.1$~GeV, increasing $\epsilon_0$
raises the unshifted baseline limit $x_{\min}^{(0)}=\epsilon_0/\lambda$ and,
through Eq.~\eqref{eq:xmin_shifted}, the two branch-specific limits, pushing
the convolutions toward the large-$x$ endpoint.  This phase-space effect overwhelms the simultaneous
decrease of $\tau$, so the scan must not be interpreted as a monotonic scan of
``TMC strength'' alone.
The physical moment ratios likewise need not vary monotonically with
$\epsilon_0$, because the scan changes $\tau$, the boundary scale
$\kappa=m_N/\epsilon_0$, and the relative importance of the moving physical-domain
boundary at the same time.

\begin{table}[h]
\centering
\caption{Combined CT18 sensitivity to $\epsilon_0$ at $Q=10$~GeV through both the trace correction and the partonic support. The column $x_{\min}^{(0)}(4.1)$ shows the unshifted no-TMC baseline limit $\epsilon_0/\lambda$ at $\sqrt{s}=4.1$~GeV; the exact TMC branches use Eq.~\eqref{eq:xmin_shifted}. The last column is therefore not an isolated scan of the trace parameter $\tau$, and no absolute-normalization comparison between different $\epsilon_0$ values is implied.}
\label{tab:epsilon0_sensitivity}
\begin{tabular}{cccccccc}
\toprule
$\epsilon_0$ [GeV] & $\tau$ & $\kappa$ & $x_{\min}^{(0)}(4.1)$ & $R_4$ & $R_6$ & $R_8$ & $\mathcal R_{\rm TMC}(4.1)$\\
\midrule
0.16 & 8.59 & 5.862 & 0.156 & 0.852 & 0.690 & 0.589 & 0.390 \\
0.25 & 3.52 & 3.752 & 0.244 & 0.863 & 0.669 & 0.538 & 0.275 \\
0.40 & 1.37 & 2.345 & 0.391 & 0.887 & 0.663 & 0.485 & 0.092 \\
0.60 & 0.61 & 1.563 & 0.586 & 0.923 & 0.718 & 0.515 & 0.001 \\
\bottomrule
\end{tabular}
\end{table}

Several independent representations provide stringent consistency checks.  The
closed Euclidean resummation reproduces the direct double series, the physical
discontinuity agrees with direct complex continuation, and the spectral,
shifted-cross-section, and physical-weight moment representations close with a maximum scaled mismatch of approximately $7.2\times10^{-6}$ across the tested PDF shapes and $n=4,6,8,10$.  The quoted PDF uncertainties are obtained
from the native uncertainty ensembles of the four sets.

\subsection{Photoproduction context}
The quantity $\sigma_{\Phi N}$ obtained above is the leading-twist OPE representation of the absorptive part of the
forward quarkonium--nucleon amplitude and is not directly the exclusive
$\gamma p\to J/\psi p$ cross section.  A photoproduction prediction requires
the vector-meson-dominance conversion used in the original OPE application
\cite{Kharzeev:1999}, a dispersive determination of the real part
\cite{Gryniuk:2016mpk}, and the measured momentum-transfer dependence entering
the conversion from a forward amplitude to an integrated exclusive cross
section \cite{GlueX2023}. Possible off-shell corrections to the
vector-meson-dominance relation are also not included. A broader construction
that matches the threshold OPE contribution to an effective high-energy
component is discussed separately in \cite{Syamtomov:photoprod}.

\begin{figure}[H]
\centering
\includegraphics[width=0.82\textwidth]{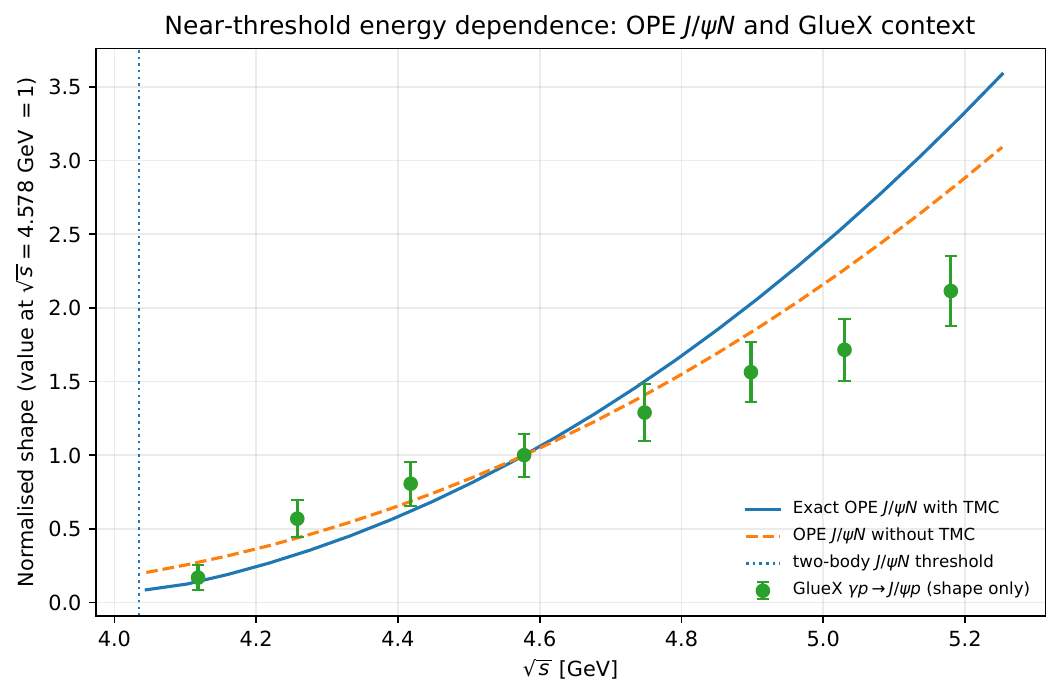}
\caption{Normalised near-threshold energy dependence of the leading-twist OPE
$J/\psi N$ spectral baseline, with and without TMC, compared with GlueX exclusive
$\gamma p\to J/\psi p$ data \cite{GlueX2023}.  The theory curves are
normalised at $\sqrt{s}=4.578$~GeV; the data normalisation at the same energy is
obtained by linear interpolation. The comparison is kinematic and shape-only:
$\sigma_{\Phi N}$ is the absorptive forward input, not a complete
photoproduction prediction, and no phenomenological agreement is inferred.}
\label{fig:photoproduction-context}
\end{figure}

Figure~\ref{fig:photoproduction-context} therefore compares only normalised
shapes.  The exact TMC and no-TMC $J/\psi N$ curves are normalised at
$\sqrt{s}=4.578$~GeV.  The GlueX normalisation at that energy is obtained by
linear interpolation of the tabulated data \cite{GlueX2023}.  No absolute
normalisation or fit is attempted. The comparison only places the localised
TMC curvature in the experimentally probed near-threshold interval and is not
used as evidence for phenomenological agreement.

\subsection{Higher-twist limitations}
The word ``exact'' in this article refers to the complete kinematic target-mass trace series within the adopted leading-twist Coulombic OPE. It does not include genuine higher-twist operators, finite-size corrections to the Coulombic quarkonium state, nonperturbative modifications of the partonic dissociation kernel, or the detailed hadronic threshold structure. Additional multipole operators and quarkonium polarisabilities may be important at $J/\psi$ threshold and require independent nonperturbative input beyond the scope of this analysis.

\section{Conclusions}
\label{sec:conclusions}

The finite nucleon mass has a pronounced but highly localised impact on the
twist-2 OPE description of $J/\psi N$ scattering.  By resumming the complete
trace series before analytic continuation, we obtained a closed physical
spectral kernel, an exact shifted-convolution spectral baseline, and a physical
moment that retains the full $y\leq1$ kinematic correlation.  This construction
removes the need for a spin or trace-order truncation and makes the connection
between the OPE and its absorptive spectral baseline explicit.

The moment ratios show a systematic hierarchy.  At $Q=10$~GeV and
$\epsilon_0=0.16$~GeV, the exact TMC/no-TMC ratio is close to $0.85$ for
$n=4$ for all four modern PDF families, but decreases to about $0.42$--$0.48$
by $n=12$ as the moments become increasingly sensitive to large $x$.  The
leading-twist spectral baseline exhibits an even more distinctive pattern.
At the strict incoming two-body endpoint the shifted representation has an
integrable inverse-velocity behaviour. This unresolved layer is a property of
the partonic/OPE construction and is not presented as a complete hadronic-threshold
prediction. In the tabulated energy range, which starts $10$~MeV above the endpoint,
the ratio is strongly suppressed and then recovers rapidly, reaching about
$0.91$ near $5$~GeV and becoming nearly unity by $10$~GeV. The effect is therefore
potentially important for resolved near-endpoint OPE applications, but it does
not persist as a large correction at higher energy. The fixed-order decomposition further shows that
this pattern is driven predominantly by the first trace term, with $j\leq2$
already tracking the exact result closely outside the first few tens of MeV
above the incoming endpoint.

Native LHAPDF grids and their complete uncertainty ensembles show that the
lowest convergent moment ratio is remarkably stable across ABMP16, MSHT20,
CT18 and NNPDF4.0.  The PDF spread grows for higher moments and in the immediate
endpoint region, where the convolution samples the least constrained large-$x$
gluon.  The PDF-input-scale dependence at fixed leading-order Coulombic
coefficients remains moderate, whereas the binding-energy parameter can
substantially alter the resolved near-endpoint baseline
because it controls both the trace parameter and the partonic support.

The result therefore provides a controlled leading-twist TMC spectral baseline. Genuine higher-twist multipole effects, finite-size corrections, and detailed hadronic threshold dynamics require independent nonperturbative input.

\section*{Acknowledgements}
The author thanks Christian Otto for stimulating discussions.

\appendix
\setcounter{table}{0}
\setcounter{figure}{0}
\section{Auxiliary analytic representations of the gluon PDFs}
\label{app:pdf_params}

All tabulated numerical moments, resolved-energy spectral baselines, uncertainty bands, and parameter scans in this work are obtained from the native LHAPDF grids and their complete uncertainty ensembles. The analytic form introduced in Eq.~\eqref{eq:pdf_param} has a much narrower purpose: it provides a smooth, compact representation of each central gluon distribution for exploratory checks and visualisation when repeated interpolation of a native grid would obscure the underlying $x$ dependence. It is not used for any tabulated or quoted quantitative result.

\begin{figure}[H]
\centering
\includegraphics[width=0.92\textwidth]{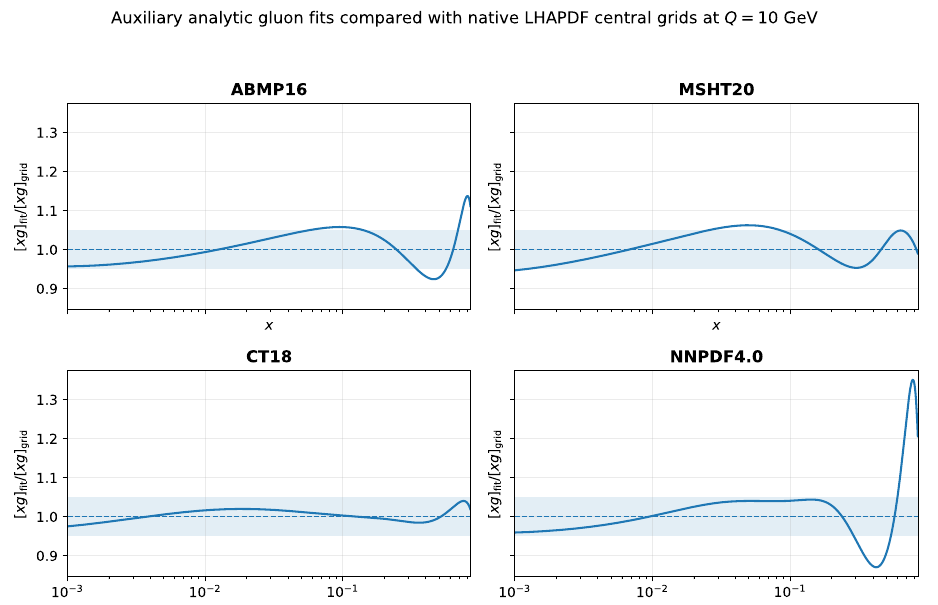}
\caption{Ratio of the auxiliary analytic representation to the native LHAPDF central gluon grid at $Q=10$~GeV.  The shaded interval marks a $\pm5\%$ deviation from the native grid.  The comparison is restricted to $x\leq0.85$, before both numerator and denominator become numerically negligible.  The fits provide a convenient smooth representation over most of this range. They are not used for any tabulated or quoted quantitative result.}
\label{fig:analytic_fit_validation}
\end{figure}

For completeness, Table~\ref{tab:analytic_pdf_params} gives the parameters of the auxiliary fits at the central scale $Q=10$~GeV.  They were obtained over the interval $10^{-5}\leq x\leq0.95$.  Figure~\ref{fig:analytic_fit_validation} compares the resulting $xg(x,Q)$ directly with the corresponding native central grids.  The comparison shows both why the fits are useful as smooth analytic representations and why the native grids remain essential for quantitative results, especially toward large $x$.

\begin{table}[H]
\centering
\caption{Parameters of the auxiliary representation $xg(x,Q)=A x^{\alpha}(1-x)^{\beta}(1+\epsilon\sqrt{x}+\gamma x)$ at $Q=10$~GeV.  The fits are used only for smooth diagnostics and visualisation; all tabulated and quoted quantitative results and uncertainty bands use native LHAPDF grids.}
\label{tab:analytic_pdf_params}
\begin{tabular}{lrrrrr}
\toprule
Set & $A$ & $\alpha$ & $\beta$ & $\epsilon$ & $\gamma$\\
\midrule
ABMP16    & 2.309 & $-0.294$ & 5.497 & $-2.050$ & 1.234\\
MSHT20    & 2.713 & $-0.261$ & 3.631 & $-2.674$ & 2.049\\
CT18      & 2.407 & $-0.284$ & 4.631 & $-2.579$ & 2.259\\
NNPDF4.0  & 2.617 & $-0.269$ & 5.911 & $-2.425$ & 2.186\\
\bottomrule
\end{tabular}
\end{table}

\setcounter{figure}{0}
\section{The $n=2$ moment and infrared sensitivity}
\label{app:n2_sensitivity}
\begin{figure}[H]
\centering
\includegraphics[width=0.88\textwidth]{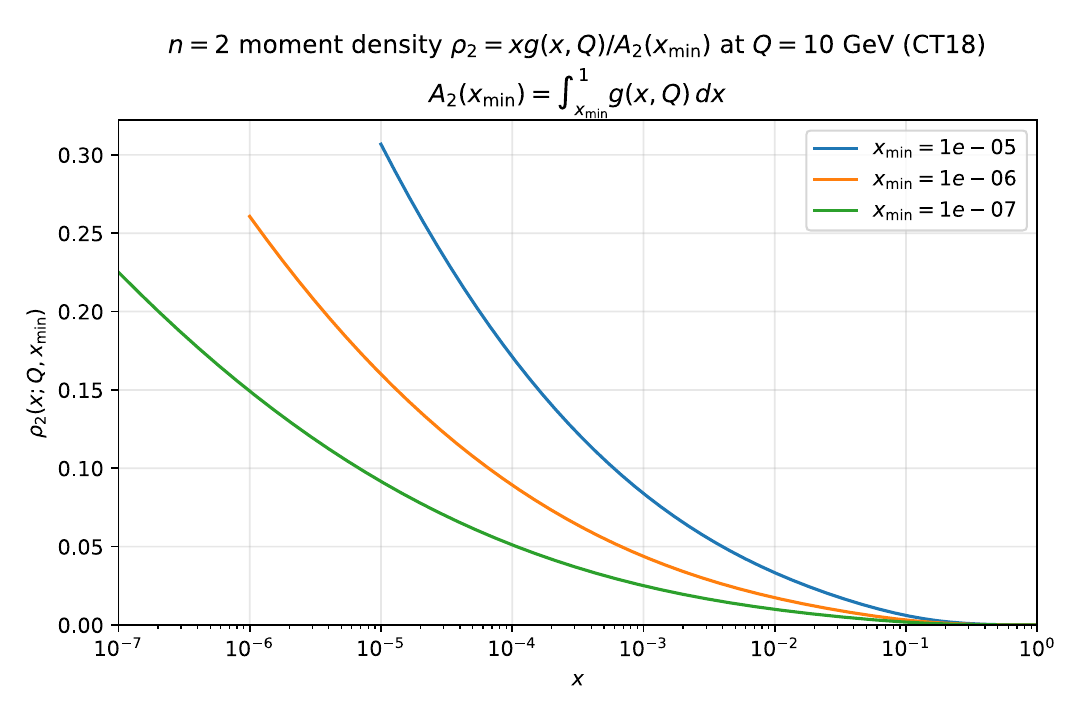}
\caption{Infrared sensitivity of the normalized $n=2$ density $\rho_2=xg(x,Q)/A_2(x_{\min})$ per $d\ln x$ for CT18 at $Q=10$~GeV.  For each curve, $A_2(x_{\min})=\int_{x_{\min}}^1g(x,Q)\,dx$ and the density integrates to unity over the displayed cutoff domain.}
\label{fig:n2_appendix}
\end{figure}

The formal Mellin moment $A_2=\int_0^1g(x,Q)\,dx$ is infrared sensitive for modern small-$x$-singular gluon distributions. It is therefore excluded from the main normalized Mellin-density diagnostics. With an explicit cutoff, the density per $d\ln x$ is
\begin{equation}
 \rho_2(x;Q,x_{\min})=\frac{xg(x,Q)}{A_2(x_{\min})},\qquad
 A_2(x_{\min})=\int_{x_{\min}}^1dx'\,g(x',Q),
\end{equation}
so that $\int_{x_{\min}}^1\rho_2(x;Q,x_{\min})\,d\ln x=1$. The plotted curves were calculated with this definition; direct numerical integration gives $1.000002$ for each of the three cutoffs shown. This infrared sensitivity does not obstruct the finite-energy cross section: the no-TMC baseline has $x_{\min}^{(0)}=\epsilon_0/\lambda$, while the exact TMC branches have the positive limits in Eq.~\eqref{eq:xmin_shifted}. Figure~\ref{fig:n2_appendix} shows explicitly how the normalized profile changes when the cutoff is lowered, in contrast to the stable convergent moments used in the main analysis.

\footnotesize
\section{Fixed-trace-order expansion of the resummed TMC kernel}
\label{app:fixed_j_expansion}

This appendix gives the independent fixed-order consistency checks used in
Secs.~\ref{sec:physical_kernel} and~\ref{sec:fixedj_spread}.  Before taking the discontinuity, the
trace-resummed generating function can be expanded as
\begin{equation}
 \mathcal F_{\rm TMC}(z,b)
 =\sum_{j=0}^{\infty}\left(-\frac{b}{z^2}\right)^j S_j(z),
 \qquad \mathcal D\equiv z\frac{d}{dz}.
 \label{eq:fixedj_series_app}
\end{equation}
The first three coefficient functions obtained independently from the finite
trace combinatorics are
\begin{align}
 S_0(z)&=F_0(z),
 \label{eq:S0_app}\\
 S_1(z)&=(\mathcal D+1)F_0(z)-d_2,
 \label{eq:S1_app}\\
 S_2(z)&=\frac12\mathcal D(\mathcal D-1)F_0(z)-d_4 z^2.
 \label{eq:S2_app}
\end{align}
Consequently, the small-$b$ expansion of the closed result
Eq.~\eqref{eq:Ftmc_exact} is
\begin{align}
 \mathcal F_{\rm TMC}(z,b)
 ={}&F_0(z)
 -\frac{b}{z^2}\left[(\mathcal D+1)F_0(z)-d_2\right]
 \nonumber\\
 &+\frac{b^2}{z^4}
 \left[\frac12\mathcal D(\mathcal D-1)F_0(z)-d_4z^2\right]
 +\mathcal O(b^3).
 \label{eq:Ftmc_fixedj_app}
\end{align}
Equivalently,
\begin{equation}
 \left.\frac{1}{j!}\frac{\partial^j\mathcal F_{\rm TMC}(z,b)}
 {\partial b^j}\right|_{b=0}
 =(-1)^j z^{-2j}S_j(z),
 \qquad j=0,1,2.
 \label{eq:Ftmc_derivative_app}
\end{equation}

The subtraction terms $d_2$ and $d_4z^2$ are entire functions of $z$ and
therefore do not contribute to the physical discontinuity.  Defining
\begin{equation}
 \phi_j(z)=\frac{1}{2i}\Disc S_j(z),
 \label{eq:phij_def_app}
\end{equation}
one obtains
\begin{equation}
 \Phi_{\rm TMC}(z,b)
 =\phi_0(z)-\frac{b}{z^2}\phi_1(z)
 +\frac{b^2}{z^4}\phi_2(z)+\mathcal O(b^3),
 \label{eq:Phi_fixedj_app}
\end{equation}
where, for $z>1$,
\begin{align}
 \phi_0(z)&=\frac{2C\pi}{3}\frac{(z-1)^{3/2}}{z^6},
 \label{eq:phi0_app}\\
 \phi_1(z)&=\frac{C\pi}{3z^6}\sqrt{z-1}\,(10-7z),
 \label{eq:phi1_app}\\
 \phi_2(z)&=\frac{C\pi}{4z^6}
 \frac{33z^2-88z+56}{\sqrt{z-1}}.
 \label{eq:phi2_app}
\end{align}
Thus the spectral kernel obeys the independent derivative test
\begin{equation}
 \left.\frac{1}{j!}\frac{\partial^j\Phi_{\rm TMC}(z,b)}
 {\partial b^j}\right|_{b=0}
 =(-1)^j z^{-2j}\phi_j(z),
 \qquad j=0,1,2.
 \label{eq:Phi_derivative_app}
\end{equation}
The first equality checks the Euclidean resummation, while the second checks
its analytic continuation and physical branch orientation.  These fixed-$j$
relations are used only as analytic and numerical consistency checks; all
main-text results employ the closed all-orders kernel
Eq.~\eqref{eq:Phi_exact}.

\section{Extended-domain factorisation and the hypergeometric moment}
\label{app:extended_hypergeom}

This appendix gives the explicit bridge between the fixed-trace expansion and
the factorised hypergeometric moment of Eq.~\eqref{eq:hypergeom_weight}. At
fixed trace order $j$, the physical change of variables
\begin{equation}
 z=\frac{\kappa x}{y},\qquad
 y=\frac{\kappa x}{z},\qquad
 dy=-\frac{\kappa x}{z^2}\,dz
 \label{eq:appD_change_variables}
\end{equation}
combines the partonic cut $z\geq1$ with the kinematic condition $y\leq1$ to
give
\begin{equation}
 y_{\max}(x)=\min(1,\kappa x),\qquad
 z_{\min}(x)=\max(1,\kappa x).
 \label{eq:appD_physical_domain}
\end{equation}
The moving lower limit is the origin of the nonfactorisation of the exact
physical moment.

The analytically extended construction replaces
$y_{\max}(x)=\min(1,\kappa x)$ by $y_{\max}(x)=\kappa x$ for all $x$ and omits
the physical phase-space restriction. This continuation into $y>1$ is an
analytic device used to expose factorisation; it is not an integration over
physical scattering states. Equivalently, $z_{\min}(x)$ is replaced
by the fixed limit $1$. The fixed-$j$ moment then separates:
\begin{align}
 M_{n,\mathrm{ext}}^{(j)}
 &=K_\sigma(-\tau)^j\kappa^{n-1}
 \int_0^1 dx\,x^{n+2j-2}g(x,Q)
 \int_1^\infty dz\,z^{-(n+2j-1)}\phi_j(z)\nonumber\\
 &=K_\sigma(-\tau)^j\kappa^{n-1}J_j(n)A_{n+2j}(Q),
 \label{eq:appD_fixedj_factorized}
\end{align}
where
\begin{equation}
 J_j(n)=\int_1^\infty dz\,z^{-(n+2j-1)}\phi_j(z).
 \label{eq:appD_Jj}
\end{equation}
For the convergent moments $n\geq4$, the termwise dispersion moments are finite. Using $\phi_j=(2i)^{-1}\Disc S_j$ and the defining series for $S_j$ in Eq.~\eqref{eq:double_trace_series}, the fixed-$j$ coefficient can therefore be obtained by matching the corresponding term proportional to $z^{n+2j-2}$. This gives the
intermediate identity
\begin{equation}
 J_j(n)=\frac{\pi}{2}\binom{n+j}{j}d_{n+2j}.
 \label{eq:appD_J_intermediate}
\end{equation}
Hence the kernel moments obey
\begin{equation}
 J_0(n)=\frac{\pi}{2}d_n,\qquad
 J_j(n)=\binom{n+j}{j}\frac{d_{n+2j}}{d_n}J_0(n).
 \label{eq:appD_J_identity}
\end{equation}
The binomial coefficient is the same combinatorial factor generated by the
$j$ trace contractions in the fixed-order expansion.
Using $K_\sigma J_0(n)=I(n)$, Eq.~\eqref{eq:appD_fixedj_factorized}
therefore becomes
\begin{equation}
 M_{n,\mathrm{ext}}^{(j)}
 =\kappa^{n-1}I(n)(-\tau)^j
 \binom{n+j}{j}\frac{d_{n+2j}}{d_n}A_{n+2j}(Q).
 \label{eq:appD_fixedj_target}
\end{equation}
Summing over $j$ and inserting
$A_{n+2j}=\int_0^1dx\,x^{n+2j-2}g(x,Q)$ gives
\begin{align}
 M_{n,\mathrm{ext}}
 &=\kappa^{n-1}I(n)\int_0^1dx\,x^{n-2}g(x,Q)
 \sum_{j=0}^\infty
 \binom{n+j}{j}\frac{d_{n+2j}}{d_n}(-\tau x^2)^j\nonumber\\
 &=\kappa^{n-1}I(n)\int_0^1dx\,x^{n-2}g(x,Q)T_n(x),
 \label{eq:appD_hypergeom_result}
\end{align}
where the series in the first line is precisely the ${}_3F_2$ function in
Eq.~\eqref{eq:hypergeom_weight}. Thus
$M_{n,\mathrm{ext}}=\kappa^{n-1}I(n)\,\widetilde A_n^{\mathrm{ext}}$,
which closes the notation between Appendix D and Sec.~2.3. This proves that the traditional
hypergeometric expression is the exact all-trace-order result for the extended domain.
It is not the exact physical $y\leq1$ moment, because restoring
Eq.~\eqref{eq:appD_physical_domain} makes the spectral lower limit depend on
$x$ and prevents the separation in Eq.~\eqref{eq:appD_fixedj_factorized}.

\end{document}